\documentclass{pnastwo}

\usepackage{graphicx} 
\usepackage{color}

\usepackage{amssymb,amsfonts,amsmath}

\contributor{Submitted to Proceedings of the National Academy of Sciences of the United States of America}
\url{www.pnas.org/cgi/doi/10.1073/pnas.0709640104}
\copyrightyear{2008}
\issuedate{Issue Date}
\volume{Volume}
\issuenumber{Issue Number}

\begin{document}


\title{A route to thermalization in the $\alpha$-Fermi-Pasta-Ulam system} 






\author{M. Onorato\affil{1}{Dip. di Fisica, Universit\`{a} di Torino
    and INFN, Sezione di Torino, Via P. Giuria, 1 - Torino, 10125,
    Italy}, L. Vozella\affil{1}{} D. Proment\affil{2}{School of
    Mathematics, University of East Anglia, Norwich Research Park,
    Norwich, NR4 7TJ, United Kingdom} \and
  Y. V. Lvov\affil{3}{Department of Mathematical Sciences, Rensselaer
    Polytechnic Institute, Troy, New York 12180, USA}}

\contributor{Submitted to Proceedings of the National Academy of
  Sciences of the United States of America}


\maketitle 

\begin{article}


\begin{abstract}
We study the original $\alpha$-Fermi-Pasta-Ulam (FPU)
system with $N=16,32$ and $64$ masses connected by a nonlinear
quadratic spring.  Our approach is based on resonant wave-wave
interaction theory, i.e. we assume that, in the weakly nonlinear
regime (the one in which Fermi was originally interested), the large
time dynamics is ruled by exact resonances. After a detailed analysis
of the $\alpha$-FPU equation of motion, we find that the first non
trivial resonances correspond to six-wave interactions.  Those are
precisely the interactions responsible for the thermalization of the
energy in the spectrum.  We predict that for small amplitude random
waves the time scale of such interactions is extremely large and
it is of the order of $1/\epsilon^8$, where $\epsilon$ is the small
parameter in the system. The wave-wave interaction theory is not based
on any threshold: equipartition is predicted for arbitrary small
nonlinearity.  Our results are supported by extensive numerical
simulations.  A key role in our finding is played by the {\it Umklapp}
(flip over) resonant interactions, typical of discrete systems.  The
thermodynamic limit is also briefly discussed.
\end{abstract}






\abbreviations{Nonlinear waves | FPU recurrence | resonant interactions}



\section{Introduction}

The Fermi-Pasta-Ulam (FPU) chains is a simple mathematical model
introduced in the fifties to study the thermal equipartition in
crystals \cite{fermi1955studies}.  The model consists of $N$ identical
masses each one connected by a nonlinear spring; the elastic force can
be expressed as a power series in the spring deformation $\Delta x$:
\begin{equation}
F=-\gamma \Delta x+\alpha \Delta x^2 +\beta \Delta x^3 +..., \label{force}
\end{equation}
where $\gamma,\alpha$ and $\beta$ are elastic, spring dependent,
constants.  The $\alpha$-FPU chain, the system studied herein,
corresponds to the case of $\alpha \ne 0$ and $\beta=0$.  Fermi, Pasta
and Ulam integrated numerically the equation of motion and conjectured
that, after many iterations, the system would exhibit a
thermalization, i.e.\! a state in which the influence of the initial
modes disappears and the system becomes random, with all modes excited
equally (equipartition of energy) on average.  Contrary to their
expectations, the system exhibited a very complicated quasi-periodic
behavior. This phenomenon has been named ``FPU recurrence'' and this
finding has spurred many great mathematical and physical discoveries
such as integrability \cite{zakharov1991integrability} and soliton
physics \cite{zabusky1965}.

More recently, very long numerical simulations have shown a clear
evidence of the phenomenon of equipartition, see for instance
\cite{benettin2013fermi} and references therein.  Yet, despite
substantial progresses on the subject
~\cite{ford1992fermi,weissert1999genesis,berman2005fermi,carati2005fermi,gallavotti2008fermi,Jack},
up to our knowledge no complete understanding of the original problem
has been achieved so far and the numerical results of the original
$\alpha$-FPU system remain largely unexplained from a theoretical
point of view.  More precisely, the physical mechanism responsible for
a first {\it metastable state} \cite{benettin2013fermi} and the
observation of equipartition for very large times have not been
understood.

In this manuscript, we study the FPU problem using an approach based
on the nonlinear interaction of weakly nonlinear dispersive waves.
Our main assumption is that the irreversible transfer of energy in the
spectrum in a weakly nonlinear system is achieved by exact resonant
wave-wave interactions. Such resonant interactions are the base for
the so called {\it wave turbulence theory}
~\cite{falkovich1992kolmogorov,nazarenko2011wave} and are responsible
for the phenomenon of thermalization.  Specifically, we will show that
in the $\alpha$-FPU system six-wave resonant interactions are
responsible for an effective irreversible transfer of energy in the
spectrum.

\section{The Model}

The equation of motion for a chain of $N$ identical particles of mass
$m$, subject to a force of the type in (\ref{force}) with $\alpha \ne
0$ and $\beta=0$, has the following form:
\begin{equation}
m\ddot
q_j=\left(q_{j+1}+q_{j-1}-2q_j\right)\left(\gamma+\alpha(q_{j+1}-q_{j-1})\right),
\label{springnl2}
\end{equation}
with $j=0,1,..,N-1$.  Here $q_j(t)$ is the displacement of the
particle $j$ from the equilibrium position. We consider periodic
boundary conditions, i.e. $q_{N}=q_{0}$.  Our approach is developed in
Fourier space and the following definitions of the direct and inverse
Discrete Fourier Transform are adopted:
\begin{equation}
Q_k=\frac{1}{N}\sum_{j=0}^{N-1} q_j e^{-i 2\pi k
  j/N},\;q_j=\sum_{k=-N/2+1}^{N/2} Q_k e^{ i 2\pi j k/N}, \label{DFT}
\end{equation}
where $k$ are discrete wavenumbers and $Q_k$ are the Fourier
amplitudes.

\subsection{Normal modes}

We then introduce the complex amplitude of a normal mode $a_k=a(k,t)$
as:
\begin{equation}
a_k=\frac{1}{\sqrt{2 \omega_k}}(P_k-i \omega_k Q_k),\label{NormalMode}
\end{equation}
where $\omega_k=\omega(k)$ is the angular frequency related to
wave-numbers as follows:
\begin{equation}
\omega_k=2 \sqrt{\frac{\gamma}{m}}|\sin(\pi k/N)|, \label{Dispersion}
\end{equation}
and $P_k$ is the momentum, $P_k=\dot Q_k$.  Substituting the above
definitions into the equation of motion (\ref{springnl2}) and
introducing the nondimensional variables
\begin{equation}
\begin{split}
a'_k=\frac{(\gamma/m)^{1/4}}{\sqrt{\sum_k \omega_k |a_k(t=0)|^2}}a_k, \;\;\;t'= \sqrt{\frac{\gamma}{m}}t,\;\;\
\omega_k'=\sqrt{\frac{m}{\gamma}}\omega_k,
\end{split}
\end{equation}
we get the
following evolution equation:
\begin{equation}
\begin{split}
&i \frac{\partial a_1}{\partial t}= \omega_1 a_1+\epsilon  \sum\limits_{k_2, k_3} V_{1,2,3}
   \bigg(
a_2a_3 \delta_{1,2+3} +2 a_2^*a_3 \delta_{1,3-2}+ \\
&+ a_2^*a_3^*
   \delta_{1,-2-3}\bigg), \label{evolu}
\end{split}
\end{equation}
where primes have been omitted for brevity and the summation on $k_2$
and $k_3$ is intended from $-N/2+1$ to $N/2$; $a_i=a(k_i,t)$,
$\delta_{i,j}=\delta_{k_i,k_j}$ is the Kronecker delta that should be
understood with ``modulus $N$'', i.e. it is also equal to one if the
argument differs by $N$. The dispersion relation becomes now
$\omega_k=2|\sin(\pi k/N)|$.
The matrix $V_{1,2,3}$ weights the transfer of energy between
 wave numbers $k_1$, $k_2$ and $k_3$ and is given by:
\begin{equation}
V_{1,2,3}=-\frac{1}{2\sqrt{2}}\frac{ \sqrt{\omega_1 \omega_2\omega_3}}
 {\mbox{sign}[\sin(\pi k_1/N)\sin(\pi k_2/N)\sin(\pi k_3/N)]}.
 \label{MatrixElement}
\end{equation}
The parameter $\epsilon$, given by
\begin{equation}
\epsilon=\frac{\alpha}{m}\left(\frac{\gamma}{m}\right)^{1/4}{\sqrt{\sum_k \omega_k |a_k(t=0)|^2}},
\end{equation}
is the only free parameter of the model. If $\epsilon=0$ the system is
linear; in the present manuscript we are interested in the weakly
nonlinear regime, i.e. $\epsilon\ll1$.
\subsection{Absence of three wave resonant interactions and the role of the  canonical transformation}
Equation (\ref{evolu}), which has a Hamiltonian structure with
canonical variables $\{ i a_k, a_k^*\}$, describes the time evolution
of the amplitudes of the normal modes of the
$\alpha$-FPU system.  It is characterized by a quadratic nonlinearity,
i.e. a three-wave interaction system.  Wave numbers $k_1$, $k_2$ and
$k_3$ are called {\it resonant} if they satisfy the following
equations:
\begin{equation} 
k_1\pm k_2 \pm k_3{\overset{N}{=}} 0 ,\;\;\;
\omega_1\pm\omega_2\pm\omega_3=0, \label{ThreeWaveResonance}
\end{equation}
where the ${\overset{N}{=}}$ sign means ``equal modulus $N$'', i.e.
wave numbers may scatter over the edge of the Brillouin zone because
of the Umklapp (flip over) scattering \cite{zuelicke2000umklapp}.
Using Prosthaphaeresis formulas one can show that it is impossible to
find non zero $k_1$, $k_2$ and $k_3$ satisfying
(\ref{ThreeWaveResonance}) with $\omega(k)$ given in
(\ref{Dispersion}).  This observation leads to a first important
consideration: the Fourier modes $a_k$ in the $\alpha$-FPU system can
be divided into {\it free} and {\it bound} modes.  To illustrate the
concept of the bound modes, we refer to the classical hydrodynamic
example of the Stokes wave in surface gravity waves, see e.g.
\cite{whitham74}.  The Stokes wave is a solution of the Euler
equations and is characterized by a primary sinusoidal wave plus
higher harmonics whose amplitudes depend on the primary wave. Those
higher harmonics are {\it bound} to the primary free sinusoidal mode
and they do not obey to the linear dispersion relation.  Cnoidal waves
and solitons are similar objects: they are characterized by a large
number of harmonics that are not free to interact with each other.  In
the light of the above comments, we then make the following statement:
the equipartition phenomenon is not to be expected for the Fourier
modes of the original variables $a(k,t)$ or $Q(k,t)$, but only for
those that are free to interact. The rest of the modes in the spectrum
do not have an independent dynamics and are phase-locked to the free
ones.  The question is then how to build a spectrum characterized only
by free modes. From a theoretical point of view, the problem can be
attacked by removing {\it via} an {\it ad hoc} canonical
transformation all interactions that are not resonant.  The
transformation inevitably generates higher order interactions which
may or may not be resonant. If those are not resonant, then a new
transformation can be applied to remove them; in principle, such
operation can be iterated up to an infinite order in nonlinearity, as
long as no resonant interactions are encountered.  In the presence of
resonant interactions, the transformation diverges because of the
classical small divisor problem \cite{arnold1963}.

For the case of $\alpha$-FPU, the following transformation from
canonical variables $\{ i a_k, a_k^*\}$ to $\{ i b_k, b_k^*\}$
\begin{equation}
\begin{split}
&a_1=b_1+\epsilon \sum\limits_{k_2, k_3} ( A_{1,2,3}^{(1)}b_2b_3
  \delta_{1,2+3}+A_{1,2,3}^{(2)} b_2^*b_3 \delta_{1,3-2}+ \\ &
  +A_{1,2,3}^{(3)}b_2^*b_3^* \delta_{1,-2-3}) + O(\epsilon^2)
 \label{transf}
\end{split}
\end{equation}
removes the triad interactions in (\ref{evolu}) and introduces higher
order nonlinearity. Here
\begin{eqnarray}
A_{1,2,3}^{(1)}&=&V_{1,2,3}/(\omega_3+\omega_2-\omega_1), \nonumber\\ 
A_{1,2,3}^{(2)}&=&2V_{1,2,3}/( \omega_3-\omega_2-\omega_1), \ \nonumber\\ 
A_{1,2,3}^{(3)}&=&V_{1,2,3}/(-\omega_3-\omega_2-\omega_1).\nonumber
\end{eqnarray}
Note that the denominators are never zero because of the non existence
of triad resonant interactions.  Higher order terms in (\ref{transf})
will be considered latter on and will involve four, five and six wave
interactions. These higher order terms will play a crucial role in the
foregoing analyses.  The procedure for calculations of such canonical
transformations is well established~\cite{krasitskii1994reduced} and
may be implemented, for example, by usage of diagrammatic technique,
as was done in \cite{Dyachenko1995233}.

In order to present a physical interpretation of the canonical
transformation, we can write it in terms of the original variable
$q_j(t)$.  Using (\ref{DFT}) and (\ref{NormalMode}), the displacement
of the masses can be written in the following form:
 \begin{equation}
q_j(t)= i  \sum_k  \left[\frac{ a_k}{\sqrt{2 \omega_k}} e^{ i 2\pi j k/N}-c.c.\right], \label{stokes0}
\end{equation}
where $c.c.$ stands for complex conjugate. We now plug (\ref{transf})
in (\ref{stokes0}) and for simplicity we assume  that the
free modes are characterized by a monochromatic wave centered in $k_0$
of the form $b(k,t)=|\bar b| \delta_{k,k_0} e^{-i(\omega_{k_0}
  t-\phi_{k_0})}$, with $\omega_{k_0}=2 |\sin(\pi k_0/N)|$, $|\bar b|$
a constant and $\phi_{k_0}$ an arbitrary phase; after some algebra,
the following result is obtained for the displacement (see also
\cite{janssen07}):
\begin{equation}
q_j(t)=A \sin(\theta)+ \epsilon B \sin(2\theta)+O(\epsilon^2), \label{stokes}
\end{equation}
with $\theta=2\pi k_0 j/N-\omega_0 t+\phi_{k_0}$, $A=-2 |\bar
b|/\sqrt{2\omega_{k_0}}$, $B=2 V_{2k_0,k_0,k_0}\sqrt{2\omega_{k_0}}
|\bar b|^2/(-4 \omega_{k_0}^2+\omega^2_{2k_0})$.  Note that $B$ is
proportional to $A^2$ and the second harmonic $2k_0$ does not
oscillate with frequency $\omega(2 k_0)$ but with $2 \omega(k_0)$,
i.e. it does not obey to the linear dispersion relation. Higher order
terms in the canonical transformation would bring higher harmonics. It
is clear that the spectrum associated to the variable $b_k$ (in the
present case, a single mode) is different from the one associated to
the variable $a_k$ or $Q_k$ (where multiple bound harmonics
appear). Equation (\ref{stokes}) is nothing but the second order
Stokes series solution of the the $\alpha$-FPU system.  The initial
stage of the $\alpha$-FPU system initialized by a single mode $k_0 $
would then be characterized by the generation of the harmonics of the
type in equation (\ref{stokes}).

\section{The reduced dynamical equation and exact resonances}

We now turn our attention to the dynamical equation that results after
the canonical transformation has been performed; the equation reads:
%
 \begin{equation}
 \begin{split}
&i \frac{\partial 
b_1}{\partial t}=  \omega_1 b_1+\epsilon^2\sum\limits_{k_2, k_3, k_4} T_{1,2,3,4}b_2^*b_3 b_4 
\delta_{1+2,3+4}+ O(\epsilon^3).
\label{evolub}
\end{split}
\end{equation}
Note that similar terms including the Kroneker deltas
$\delta_{1,3+4+2},\delta_{1,-3-4-2},\delta_{1,4-3-2}$ should also
appear in equation (\ref{evolub}) as a result of the transformation
(\ref{transf}); however, those terms are not resonant and can be
removed by higher order terms in the transformation.  The matrix
$T_{1,2,3,4}$ has an articulated analytical form which depends on
$V_{1,2,3}$ and is given for example in \cite{Dyachenko1995233} or
\cite{janssen2004interaction}.  Due to the Hamiltonian structure of
the original system, $T_{1,2,3,4}$ has the following symmetries
$T_{1,2,3,4}=T_{2,1,3,4}=T_{3,4,1,2}$. We underline that the same 
equation but with a different 
 matrix $T_{1,2,3,4}$ can be obtained  directly for the so called $\beta$-FPU 
or  for the $\alpha+\beta$ -FPU chains. 
If higher order terms are neglected, the equation admits 
 a Birkhoff normal form \cite{henrici2008results} (see
  also \cite{rink2006proof}).
Equation (\ref{evolub}), with different linear dispersion relation,
different matrix $T_{1,2,3,4}$ and with integrals instead of sums, is
the equivalent of the {\it Zakharov equation} for one directional water waves
\cite{zakharov68}.

Equation (\ref{evolub}) describes the {\it
  reduced} $\alpha$-FPU model where three-wave interactions have been
removed by the canonical transformation (\ref{transf}).  The resonant
interactions associated with equation (\ref{evolub}) are described by
the following 4-wave resonant conditions:
\begin{equation}
k_1+ k_2 - k_3 -k_4{\overset{N}{=}} 0,\;\;\; \omega_1+\omega_2-\omega_3
-\omega_4=0. \label{resonant}
\end{equation}
Solutions for $k_i\in\mathbb{Z}$ and
  $N=16$ or $32$ or $64$ particles, like in the original FPU problem, are reported  below. 
\\ {\it Trivial solutions}.  Trivial resonances are obtained when all
wave numbers are the same or when:
\begin{equation}
\begin{split}
 k_1=k_3, \; k_2=k_4,\label{Trivial}
\end{split}
\end{equation}
with permutations of 3 and 4.  These trivial solutions are responsible
for a nonlinear frequency shift and do not contribute to the energy
transfer between modes, as discussed for example in
\cite{nazarenko2011wave}. \\
{\it Nontrivial solutions}. Nontrivial resonances exist and are the
result of the following scattering process: when three waves interact
to generate a fourth one, it can happen that the latter is
characterized by a $ k \not\in [-N/2+1, ..., N/2]$,
i.e. outside the Brillouin zone.
The system flips back this energy into a mode
contained in the domain; as mentioned, this is known as Umklapp
scattering process
\cite{zuelicke2000umklapp,gershgorin2007interactions}.  These resonant
modes have the following structure:
\begin{eqnarray}
(k_1,k_2,-k_1,-k_2),
\label{FourWaveResonances}
\end{eqnarray}
with $ k_1 + k_2 = m N/2$ and $m = 0, \pm 1, \pm 2.$
It is instructive to give an example: let us consider a chain of
$N$=32 masses; the maximum wave-number accessible is then
$k_{max}=16$. One of the quadruplets satisfying the resonant condition
(\ref{resonant}) is $k_1=2$, $k_2=14$, $k_3=-14$, $k_4=30$. The first
three wave-numbers are contained in the domain, while $k_4$ is
outside. The system will interpret $k_4=30$ as $k_4\rightarrow
k_4-N=-2$.

In order to account for an effective energy mixing, the quadruplets
(four modes satisfying the resonant conditions) should be interconnected,
i.e.  single wave numbers should belong to different quartets.
However, a careful and straightforward analysis of equation
(\ref{FourWaveResonances}) reveals that resonant quartets are not
interconnected, that is, all of the quartets are isolated: if one puts
energy into one of the quartets, and only weakly nonlinear resonant
interactions are allowed, the energy will remain in the quartet.  The
above results has the important consequence that the four-wave
interactions in the $\alpha$-FPU model cannot possibly lead to
equipartition of energy.

An efficient mixing mechanism should be searched by extending to
higher order the canonical transformation.  Five-wave interactions are
non resonant and can be removed by an appropriate choice of the higher
order terms in (\ref{transf}). The resulting evolution equation for
$b_1$ is a refinement of equation (\ref{evolub}) and it has the form:
\begin{equation}
 \begin{split}
&i \frac{\partial 
b_1}{\partial t}=  \omega_1 b_1+
\epsilon^2\sum T_{1,2,3,4}b_{2}^*b_3 b_{4}+
\\
&+\epsilon^4\sum\limits W_{1,2,3}^{4,5,6}b_2^*b_3^* b_4b_5 b_6 
\delta_{1+2+3,4+5+6}+O(\epsilon^5).
\label{evolub6} 
\end{split}
\end{equation}
The explicit form of the matrix $W_{1,2,3}^{4,5,6}$, see for details
\cite{nazarenko2011wave,dyachenko2013nonintegrability}, is not
necessary for the discussion of our results. Note that the four-wave
interactions cannot be removed because, even though they do not
contribute to the spreading of energy in the spectrum, they are
resonant and a canonical transformation suitable for removing those
modes would diverge.

For $N=16$ or $32$ or $64$ we have found that solutions of the
following six-wave resonant conditions
\begin{equation}
\begin{split}
k_1+k_2+k_3-k_4-k_5-k_6&{\overset{N}{=}} 0,
\\ \omega_1+\omega_2+\omega_3-\omega_4-\omega_5-\omega_6&=0,
\label{SIX}
\end{split}
\end{equation}
exist for $k_i\in\mathbb{Z}$ and we report them below.  

{\it Trivial
  resonances}. Trivial resonances are obtained when all six wave
numbers are the same or when
\begin{equation}
 k_1=k_4,\; k_2=k_5,\;k_3=k_6, \label{SixWaveTrivial}
\end{equation}
with all permutations of indices 4,5,6. These trivial solutions are
responsible only for a nonlinear frequency shift. \\ {\it Nontrivial
  symmetric resonances}. These resonances are over the edge of the
Brillouin zone and are given by:
\begin{equation}
(k_1,k_2,k_3,-k_1,-k_2,-k_3),\label{SixWaveResonances1}
\end{equation}
 with $ k_1 + k_2 +k_3 = m N/2$ and
 $m = 0, \pm 1, \pm 2,...$ \, .\\
{\it Nontrivial quasi-symmetric resonances}. These resonances over the
edge of the Brillouin zone are characterized by one repeated wave
number:
\begin{eqnarray}
(k_1,k_2,k_3,-k_1,-k_2,k_3), \label{SixWaveResonances2}
\end{eqnarray}
with $k_1 + k_2 = m N/2$, $m= 0, \pm 1, \pm 2,...$ and all
permutations of the indices. We have not found any other integer
solutions to the resonant condition (\ref{SIX}) for $N=16, 32, 64$,
other than those in (\ref{SixWaveTrivial}), (\ref{SixWaveResonances1})
and (\ref{SixWaveResonances2}).

Those resonant sextuplets are interconnected, therefore they represent
an efficient mechanism of spreading energy in the spectrum.  Because
of the existence of these exact resonant process, we expect the
equipartition to take place.

\section{Estimation of the equipartition time scale}

Thermal equilibrium is a statistical concept; therefore, an equation for the
time evolution of the average spectral energy density is required.  
Within the {\it wave turbulence theory}, such equation is
called {\it kinetic equation}.
There are many techniques for deriving it that 
are object of intensive studies.  We are mainly interested in the estimation of the time scale of 
 equipartition; therefore, only the key steps are here considered 
(details can be found in
\cite{falkovich1992kolmogorov}).  We introduce the wave action,
$n_k=n(k,t)$, defined as $\langle b_{1}b_{2}^*
\rangle=n_1\delta_{1,2}$, where the brackets indicate ensemble
averages and the $\delta_{1,2}$ is the Kronecker delta, the latter
arising from the assumption of homogeneity of the wave field. In order
to derive the evolution equation for $n_1$, equation (\ref{evolub6})
is multiplied by $b_1^*$ and the ensemble averages are taken.  The
time evolution of the $n_1$ depends on the six-order correlator
$\langle b_1^*b_2^*b_3^*b_4b_5b_6\rangle$ whose time evolution will be
a function of higher order correlators; this is the classical BBGKY
hierarchy problem.  The time evolution of the six-order correlator
turns out to be proportional to $\epsilon^4 W_{1,2,3}^{4,5,6}$.
Indeed, assuming that the waves obey Gaussian statistics, we decompose the
higher order correlators into products of second order correlators
and, after taking the large time limit, the six order correlator may
be obtained explicitly. The result is then plug into the equation for
the time evolution of $n_k$, resulting in a collision integral
proportional to $(\epsilon^4 W_{1,2,3}^{4,5,6})^2$ (see for example
equation (6.89) in \cite{nazarenko2011wave}).  Consequently, the time
evolution of the spectral energy density in the $\alpha$-FPU problem
is proportional to $1/\epsilon^8$.  This is the main theoretical
result of our work which, as it will be shown, is supported by
numerical simulations.

\section{Relation to the Toda Lattice}

Before showing our numerical simulation results, it is instructive to
make a connection between the $\alpha$-FPU and the integrable Toda
system \cite{toda1967vibration} (see also \cite{ferguson1982Nonlinear}
for a normal mode approach to the problem).  In
\cite{benettin2013fermi} it has been shown that the $\alpha$-FPU can
be seen as a perturbation of the integrable Toda lattice, see also
\cite{casetti1997fermi}. For the sake of clarity, we report this
argument.  We consider the general Hamiltonian for a discrete
lattice
\begin{equation}
H(p,q)=\frac{1}{2}\sum_{j=1}^{N}p_j^2+\sum_{i=1}^{N} V(q_{j+1}-q_j);
\end{equation}
the potential $ V $ for the $\alpha$-FPU case is given by
\begin{equation}
V(r)=\frac{r^2}{2}+\varepsilon \frac{r^3}{3} \, . \label{fpupot} 
\end{equation}
For the Toda lattice instead $ V $ results in
\begin{equation}
V(r)=V_0(e^{\lambda r}-1-\lambda r),
\end{equation}
with $V_0$ and $\lambda$ free parameters. For the particular choice of
$V_0=1/(4\varepsilon^2)$ and $\lambda=2\varepsilon$, upon Taylor
expanding the exponential in the Toda lattice for small $\varepsilon$,
we obtain:
\begin{equation}
V(r)=\frac{r^2}{2}+\varepsilon\frac{r^3}{3}+\varepsilon^2\frac{
  r^4}{6}+\dots \ ,\label{todaexppot}
\end{equation}
that shows the $\alpha$-FPU coincides with the Toda Lattice up to the
order of $\epsilon$, see eqs. (\ref{fpupot}) and (\ref{todaexppot}).

Having made this introduction, using our approach based on resonant
interactions and, following the fundamental work by Zakharov and
Shulmann \cite{zakharov1988additional}, we are able to discern between
integrable and nonintegrable dynamics.  We underline that the Toda
lattice has the same linear dispersion relation as the $\alpha$-FPU
and, therefore, the possible resonant manifolds, being based on linear
frequencies, are exactly the same.  How then is it possible to explain that the
Toda lattice never thermalizes while the FPU does?  For the Toda
lattice, the same canonical transformation as in (\ref{transf}) can be
performed and the three wave interactions be removed (those correspond
to the term proportional to $\epsilon$ in the potential in equation
(\ref{todaexppot})).  The equation of motion in Fourier space,
neglecting higher order terms, becomes the same as the one in equation
(\ref{evolub}), with the only difference that the matrix $T_{1,2,3,4}$
is now modified due to the existence of the term proportional to
$\varepsilon^2$ in the potential (\ref{todaexppot}); such term is
absent in the potential of the $\alpha$-FPU.  A straightforward but
lengthy calculation shows that the matrix $T_{1,2,3,4}$ for the Toda
lattice is identically zero, once calculated on the resonant manifold.
Indeed, this result has very profound origins and is based on the
fundamental work by Zakharov and Shulmann
\cite{zakharov1988additional}, where it is shown that for an
integrable system, either there are no resonances or all the scattering matrices
are zero at all orders on the resonant manifold.  In principle an
infinite order canonical transformation would linearize an integrable
system.  Thus, integrable systems are characterized by trivial
scattering processes and a pure thermalization is never reached.  The
initial evolution of the spectrum observed in computations of the Toda
lattice, see for instance \cite{benettin2013fermi}, is due to non
resonant interactions that are contained in the canonical
transformation.  On the other side, for nonintegrable systems such as
the $\alpha$-FPU non trivial resonant interactions with non-zero
scattering matrix exist and thermalization can be observed.

\section{Numerical Simulations} \label{numsim}
\begin{figure}
\includegraphics[width=0.8\linewidth]{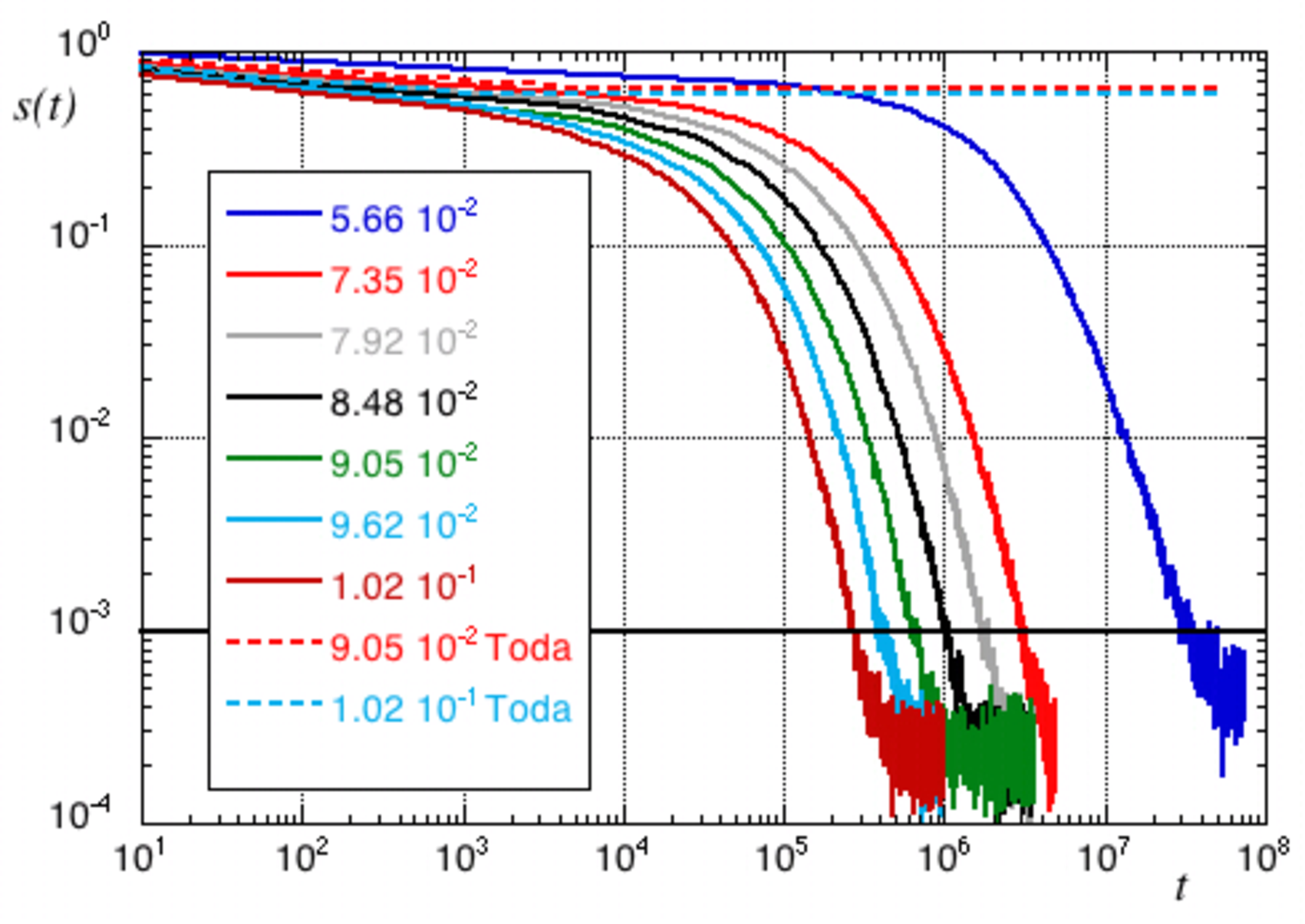}
\caption{Entropy $s(t)$ as a function of time for different simulations of the $\alpha$-FPU system
characterized by different
values of $\epsilon$. In the plot two simulations of the Toda lattice are also presented.
A horizontal line at $s=0.001$ is also included as a threshold for estimating 
the equipartition time.}
 \label{entropy_unc}
\end{figure}

A main result of this study is that the $\alpha$-FPU model should 
reach thermal equilibrium on a time scale of 
$1/\epsilon^8$ for arbitrary small nonlinearity.
Therefore, we use numerical simulations to support our theoretical
finding.  We integrate in time equation (\ref{springnl2}) with $N=32$
particles and with $m=\gamma=1$ by using the sixth order symplectic
integrator scheme described in \cite{yoshida1990construction}.
Different values of $\epsilon$ have been considered between $0.0566$ and
$0.11$; due to the slow time needed to reach thermalization,
computations become soon prohibitive for smaller values of $\epsilon$.

We emphasize that the thermal equilibrium is a statistical concept,
therefore averages should be taken to observe it.  It may be possible
to average over time, as in many previous
simulations of the FPU model.  We find such approach to be
 problematic, because often the time window used is of the same
order of the characteristic time to reach equipartition of energy.  We
have chosen to perform ensemble averaging, typically over 1000
realizations (some convergence tests have also been made over 2000
ensembles).  Two types of initial conditions have been considered: in
the first one, we have initialized only the modes $k=\pm1$; in the second
one, initial conditions are characterized by constant energy only over the
modes $k=\pm 1, \pm 2, \pm 3, \pm 4, \pm 5$.
Different random phases are then applied to the Fourier amplitudes for
each realization.

As in \cite{livi1985equipartition}, we have introduced, as an
indicator of thermalization, the following entropy:
\begin{equation}
s(t)=\sum\limits_{k} f_k \log f_k 
\end{equation}
with 
\begin{equation}
f_k = \frac{N-1}{E_{tot}}\omega_k \langle |a_k|^2 \rangle,\;\;\; E_{tot}=
\sum\limits_{k}{\omega_k \langle |a_k|^2 \rangle} 
\end{equation}
and $\langle...\rangle$ defines the average over the realizations. We
have used in the definition $N-1$ instead of $N$, because, with
periodic boundary conditions, the modes that thermalize are $N-1$ and
not $N$ (the first mode $k=0$ is not involved in the dynamics). For a
thermalized spectrum, the value of the entropy is theoretically 0.
Through our numerical simulations, we have reached a minimum value of
$s$ very close to $10^{-4}$.  In figure 1 we show the evolution of the
entropy for different values of  $\epsilon$. As one
can observe, in the large time limit, the entropy reaches very small values.
 Two typical simulations of the Toda
lattice are also included in the figure and show that, as expected, no
thermalization is reached.
Just as an example, we show in Figure 2 the energy spectrum, defined
as $E(k)=\omega_k \langle |a_k|^2 \rangle$, at different time steps
for the simulation with $\epsilon=0.0848$. The spectrum is normalized
by $(N-1)/E_{tot}$ in such a way that, once thermalization has been
reached, all values of the energy are around 1.
\begin{figure}
\includegraphics[width=0.8\linewidth]{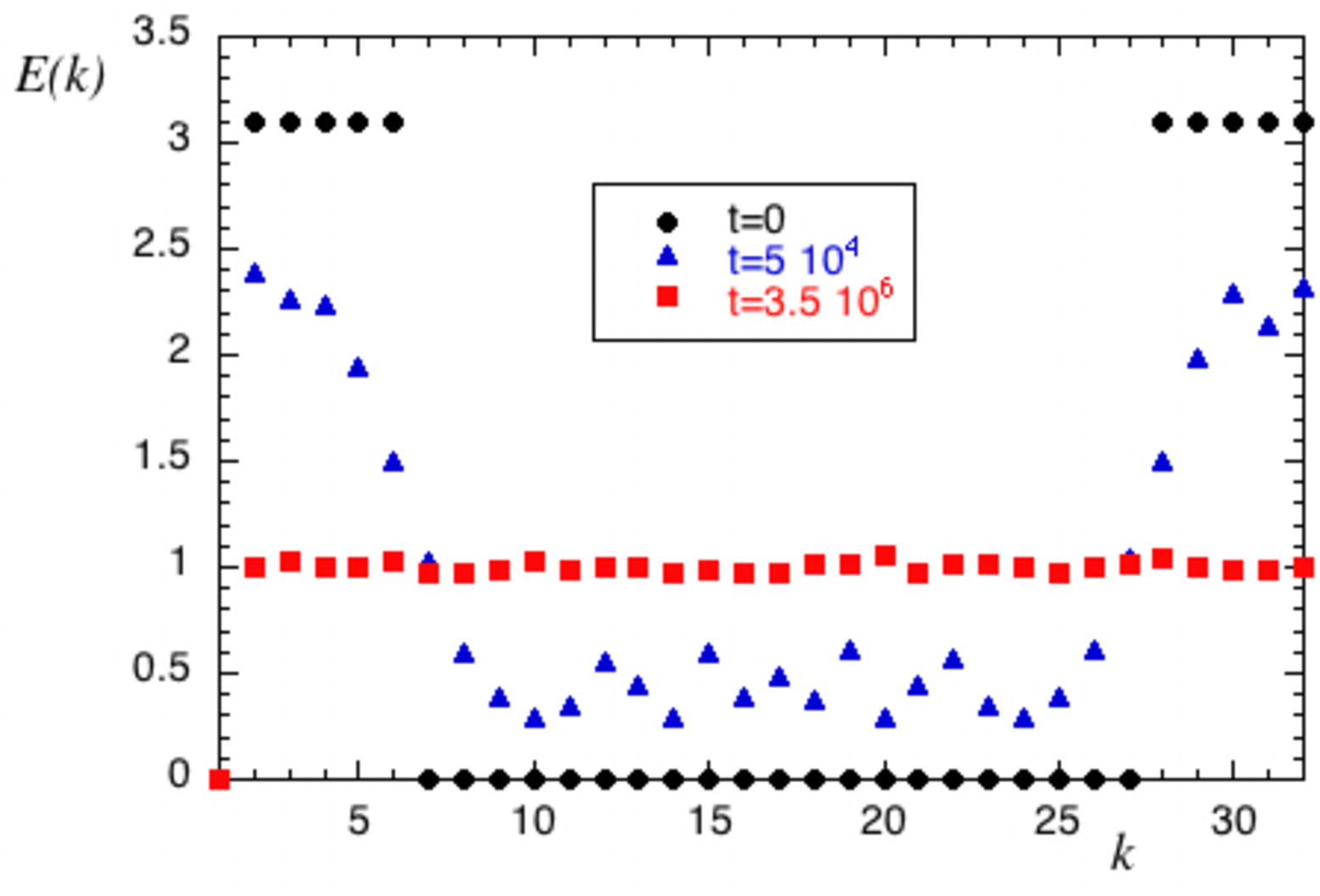}
\caption{Energy spectrum at different time steps from numerical simulations 
of the $\alpha$-FPU system. Black
  dots correspond to the initial condition; blue triangles correspond 
  to an intermediate stage and red squares to the final
  thermalized spectrum. Note
  that energy is presented in linear scale.}
 \label{spectrum}
\end{figure}

In order to verify the expected time scaling, we introduced an entropy
threshold $ s_{thr} $ to estimate the time it takes for the system to
reach thermodynamic equilibrium.  Specifically, we have defined
$t_{eq}$ as the time in which the entropy $s$ reaches the value of $
s_{thr}=0.001 $ (see an horizontal line at $s=0.001$ in Figure 1).  We
present in Figure 3 the log-log plot of this time $t_{eq}$ as a
function of $\epsilon$ for the two types of simulations
considered. Figure 3 also shows the straight line with slope -8. All
the points are pretty much aligned with this straight line.  This
numerical result is consistent with our analytic prediction that time
$t_{eq}$ is proportional to $\epsilon^{-8}$.
\begin{figure}
\includegraphics[width=0.7\linewidth]{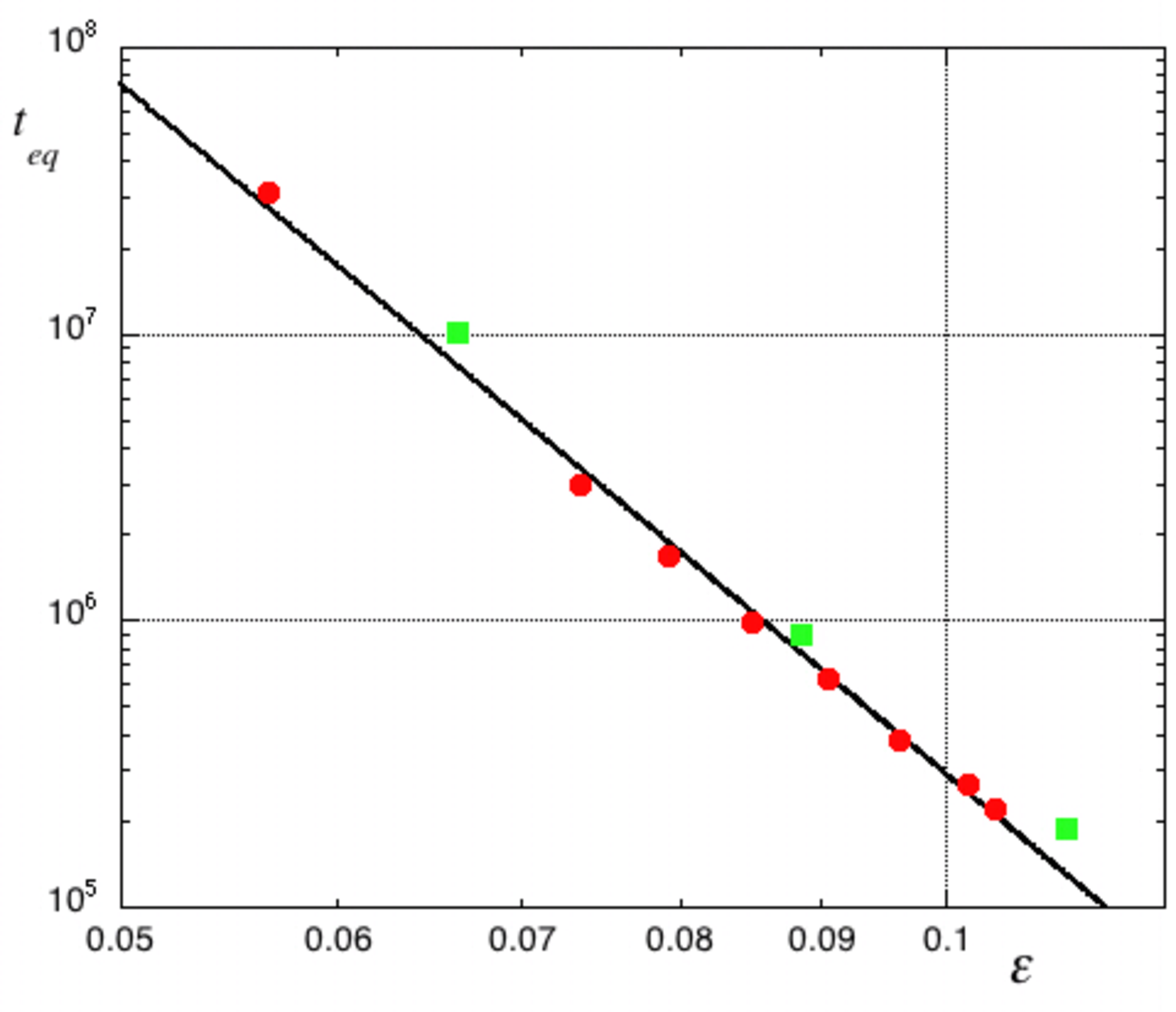}
\caption{Equilibrium time $t_{eq}$ as a function of $\epsilon$ in
  Log-Log coordinates.  Red dots represent numerical experiments 
  of the $\alpha$-FPU system with
  broad band initial conditions, i.e. modes $k=\pm
  1,\pm 2\pm3,\pm 4,\pm 5$ have been initially perturbed;
   green dots represent narrow band
  initial conditions, i.e. modes  $k=\pm 1$ have been initially perturbed.
    The straight line corresponds to power law of the type $1/\epsilon^{8}$}
 \label{scaling}
\end{figure}
The last check on the validity of the theory, free of an arbitrary
threshold, is made by rescaling the time evolution of the entropy: 
in Figure 4 we show the evolution of the entropy as a function of 
$\epsilon^8 t$ for different values of $\epsilon$. 
As predicted by the theory, the curves seem to collapse to a single one.

\begin{figure}
\includegraphics[width=0.8\linewidth]{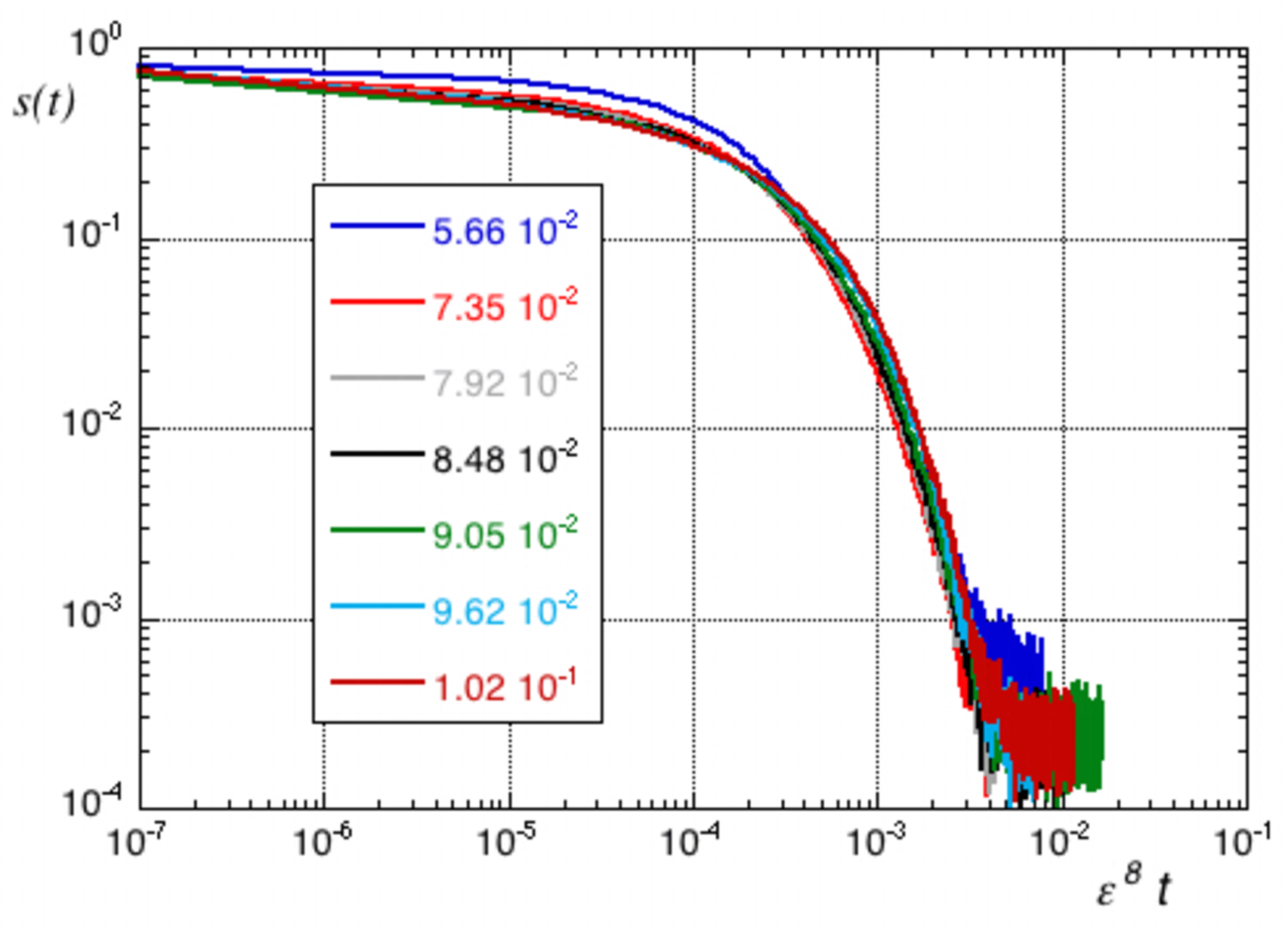}
\caption{Entropy $s(t)$ as a function of $\epsilon^8 t$ for different
  simulations  of the $\alpha$-FPU system 
  characterized by different values of $\epsilon$.}
 \label{entropy_coll}
\end{figure}

\section{A note on the thermodynamic limit}
In statistical mechanics one is usually interested in the
thermodynamic limit.  Assuming that the length of the chain is $L$ and
the spacing between masses is $\Delta x$, we let $N \rightarrow
\infty$ and $L\rightarrow \infty$ in such a way that $\Delta
x=L/N=const$. Wave numbers in Fourier space become dense, $\Delta k=2
\pi/L\rightarrow 0$. The dispersion relation now becomes
$\omega(\kappa)=2\left| \sin(\kappa/2) \right|$, where we have 
set  $\Delta x=1$ and $\kappa=k \Delta k$. Assuming that
$\kappa\in\mathbb{R}$, the same approach based on resonant
interactions can now be performed. It turns out that four wave,
non-isolated, resonant interactions exist. An example is provided by
the  following two connected quartets:
\begin{equation}
\begin{split}
(-0.05710747907971836...,1.604305030276316...,1/2,\pi/3)\nonumber\\
  (-0.12747695473542747...,2.198273281530324...,1/2,\pi/2),
\nonumber
\end{split}
\end{equation}
that can be found numerically using the resonant conditions
(\ref{SIX}).  The existence of interconnected resonant quartets
implies that in the thermodynamic limit the equipartition may be
achieved by resonant four wave interactions; in this case the system
is completely described by the traditional wave turbulence four-wave
kinetic equation~\cite{falkovich1992kolmogorov}. The time scale of the
resonant four wave kinetic equation is given by the $1/\epsilon^{4}$, 
 much shorter than $1/\epsilon^{8}$, i.e. the time scale of 
equipartition for a system of  $N$=16,32,64 masses.

\section{A short discussion  on other possible scenarios}

The FPU system has been the subject of many studies and a presentation
of all the different points of view developed in sixty years is merely
impossible. However, here we briefly present some routes to
equipartition that are accepted nowadays in the literature.

In the pioneering work of Izrailev and Chirikov \cite{izrailev1966},
the idea that an initial energy larger than a critical value is needed
in order to reach thermalization was put forward. Such concept is
based on the fact that the nonlinearity changes the linear dispersion
relation and, consequently, the resonant condition in frequency is
then modified. When the nonlinearity becomes large enough, a mechanism
of ``overlap of frequencies'' may take place. Such phenomenon lead to
the introduction of the so called ``stochastic threshold''. In the
late sixties, not everybody shared such an idea; indeed, for example
in 1970 Ford and Lunsford \cite{ford1970stochastic} insisted on the
fact that mixing could be observed also in the limit as the
nonlinearity goes to zero.

In favor of the existence of a threshold, a large number of papers
have been written and different scenarios have been presented, see for
example \cite{livi1985equipartition}.  A very interesting picture has
been presented in \cite{cretegny1998localization}: the authors
considered the $\beta$-FPU system with initial conditions
characterized by the highest mode (also known as the $\pi$-mode). They
showed that, above an energy threshold which can be computed
analytically, the $\pi$ mode is modulationally unstable and give rise
to localized chaotic structures (breathers). They related the lifetime
of the chaotic breathers to the time necessary for the system to reach
equipartition.  This interesting scenario cannot be directly
applied to the $\alpha$-FPU; the reason is that a straightforward
calculation shows that a single mode in the $\alpha$-FPU is
modulationally stable. This does not imply that in the $\alpha$-FPU
model localized coherent structure do not exist.  Indeed, being the
system close to the Korteweg de Vries equation, solitary waves may be
excited, if the initial energy is sufficiently large.  However, our
main finding is that such strong nonlinearity is not needed to reach
equipartition. Our explanation is based only on resonant interactions
and, as a result, equipartition can take place for arbitrary small
nonlinearity, as confirmed by numerical simulations.

We mention once more that our analyses are based on $N$=$16$ or $32$
or $64$ masses, as the original simulations of Fermi, Pasta and Ulam;
by changing the number of masses the solution to the resonant
conditions may change; therefore, each case should be treated
separately and possibly different scenarios may appear, as for example
the thermodynamic limit described above.

\section{Conclusion}
$i)$ Resonant triads are forbidden; this implies that, on a short time
scale, three-wave interaction will generate a reversible dynamics.
This is what has been observed originally by Fermi, Pasta and Ulam and
what is known as {\it metastable state} (see for example
\cite{benettin2013fermi}). \\
$ii)$ A suitable canonical transformation allows us to look at higher
order interactions in the system which are responsible for longer time
scale dynamics.\\
$iii)$ Four-wave resonant interactions exist;  however, we have shown that
for $N$=16,32,64 each resonant quartet is isolated, preventing the
full spread of the energy across the spectrum and thermalization.
\\
$iv)$ Six wave interactions lead to irreversible energy mixing.\\
$v)$ The time scale of equipartition in a weakly nonlinear 
random system described by $\alpha$-FPU system  is
$1/\epsilon^8$.  The result is consistent with our numerical
simulations.\\
$vi)$ In the thermodynamic limit, non-isolated resonant quartets exits and 
the time scale of equipartition is $1/\epsilon^4$.

\begin{acknowledgments}
M.O. was supported by ONR Grant No. 214 N000141010991 and
by MIUR Grant PRIN 2012BFNWZ2. M. Bustamante, M. Cencini, F. De Lillo, S. Ruffo and B. Giulinico are acknowledged  for discussion. Y.L. was supported by ONR Grant No. N00014-12-1-0280.
\end{acknowledgments}
 \bibliography{references}

\begin{thebibliography}{26}

\bibitem{fermi1955studies}
E. Fermi, J. Pasta, and S. Ulam. Studies of nonlinear problems. No. LA 1940. I, Los Alamos Scientific Laboratory Report No. LA-1940, 1955.

\bibitem{zakharov1991integrability}
Zakharov, V. E., and Schulman, E. I. (1991). Integrability of nonlinear systems and perturbation theory. In What Is Integrability? (pp. 185-250). Springer Berlin Heidelberg.

\bibitem{zabusky1965}
Zabusky, N. J., and Kruskal, M. D. (1965). Interaction of" Solitrons" in a Collisionless Plasma and the Recurrence of Initial State. Princeton University Plasma Physics Laboratory.

\bibitem{benettin2013fermi}
Benettin, G., Christodoulidi, H., and Ponno, A. (2013). The Fermi-Pasta-Ulam Problem and Its Underlying Integrable Dynamics. Journal of Statistical Physics, 1-18.

\bibitem{ford1992fermi}
Ford, J. (1992). The Fermi-Pasta-Ulam problem: paradox turns discovery. Physics Reports, 213(5), 271-310.

\bibitem{weissert1999genesis}
Weissert, T. P. (1999). The genesis of simulation in dynamics: pursuing the Fermi-Pasta-Ulam problem. Springer-Verlag New York, Inc..

\bibitem{berman2005fermi}
Berman, G. P., and Izrailev, F. M. (2005). The Fermi-Pasta-Ulam problem: fifty years of progress. Chaos (Woodbury, NY), 15(1), 15104.

\bibitem{carati2005fermi}
Carati, A., Galgani, L., and Giorgilli, A. (2005). The FermiÐPastaÐUlam problem as a challenge for the foundations of physics. Chaos: An Interdisciplinary Journal of Nonlinear Science, 15(1), 015105-015105.

\bibitem{gallavotti2008fermi}
Gallavotti, G. (Ed.). (2008). The Fermi-Pasta-Ulam problem: a status report (Vol. 728). Springer.


\bibitem{Jack} Jackson, E. A.(1978). Nonlinearity and irreversibility
  in lattice dynamics.  Rocky Mountain J. Math, 8,27-196.

\bibitem{falkovich1992kolmogorov}
Zakharov, V. E., L'vov, V. S., and  Falkovich, G. (1992). Kolmogorov spectra of turbulence 1. Wave turbulence. Kolmogorov spectra of turbulence 1. Wave turbulence., by Zakharov, VE; L'vov, VS; Falkovich, G.. Springer, Berlin (Germany), 1992, 275 p., ISBN 3-540-54533-6, 1.
 
\bibitem{nazarenko2011wave}
Nazarenko, S. (2011). Wave turbulence (Vol. 825). Springer.


\bibitem{zuelicke2000umklapp}
Papa, E., and MacDonald, A. H. (2005). Edge state tunneling in a split Hall bar model. Physical Review B, 72(4), 045324.


\bibitem{whitham74}
Whitham, G. B. (2011). Linear and nonlinear waves (Vol. 42). John Wiley and Sons.


\bibitem{arnold1963}
Arnol'd, V. I. (1963). Small denominators and problems of stability of motion in classical and celestial mechanics. Russian Mathematical Surveys, 18(6), 85-191.


\bibitem{krasitskii1994reduced}
Krasitskii, V. P. (1994). On reduced equations in the Hamiltonian theory of weakly nonlinear surface waves. Journal of Fluid Mechanics, 272(1).

\bibitem{Dyachenko1995233}
Dyachenko, A. I., Lvov, Y. V., and Zakharov, V. E. (1995). Five-wave interaction on the surface of deep fluid. Physica D: Nonlinear Phenomena, 87(1), 233-261.


\bibitem{janssen2004interaction}
Janssen, P. (2004). The interaction of ocean waves and wind. Cambridge University Press

\bibitem{janssen07}
Janssen, P. A., and Onorato, M. (2007). The intermediate water depth limit of the Zakharov equation and consequences for wave prediction. Journal of Physical Oceanography, 37(10), 2389-2400.

\bibitem{zakharov68}
Zakharov, V. E. (1968). Stability of periodic waves of finite amplitude on the surface of a deep fluid. Journal of Applied Mechanics and Technical Physics, 9(2), 190-194.

\bibitem{henrici2008results} Henrici, A., and  Kappeler T.. "Results on normal forms for FPU chains." Communications in Mathematical Physics 278.1 (2008): 145-177.

\bibitem{rink2006proof} Rink, B. (2006). Proof of Nishida's conjecture on anharmonic lattices. Communications in mathematical physics, 261(3), 613-627.

\bibitem{gershgorin2007interactions}
Gershgorin, B., Lvov, Y. V., and Cai, D. (2007). Interactions of renormalized waves in thermalized Fermi-Pasta-Ulam chains. Physical Review E, 75(4), 046603.

\bibitem{dyachenko2013nonintegrability}
Dyachenko, A. I., Kachulin, D. I., and Zakharov, V. E. E. (2013). On the nonintegrability of the free surface hydrodynamics. JETP letters, 98(1), 43-47.

\bibitem{toda1967vibration}
Toda, M. (1967). Vibration of a chain with nonlinear interaction. Journal of the Physical Society of Japan, 22(2), 431-436.

\bibitem{ferguson1982Nonlinear}
Ferguson Jr, W. E., Flaschka, H., and McLaughlin, D. W. (1982). Nonlinear normal modes for the Toda chain. Journal of Computational Physics, 45(2), 157-209.

\bibitem{casetti1997fermi}
Casetti, L., Cerruti-Sola, M., Pettini, M., and Cohen, E. G. D. (1997). The Fermi-Pasta-Ulam problem revisited: stochasticity thresholds in nonlinear Hamiltonian systems. Physical Review E, 55(6), 6566.

\bibitem{zakharov1988additional}
Zakharov, V. E., and Schulman, E. I. (1988). On additional motion invariants of classical Hamiltonian wave systems. Physica D: Nonlinear Phenomena, 29(3), 283-320.

\bibitem{ponno2011two}
  Ponno, A., Christodoulidi, H., Skokos, C., and Flach, S. (2011). The two-stage dynamics in the Fermi-Pasta-Ulam problem: from regular to diffusive behavior. Chaos: An Interdisciplinary Journal of Nonlinear Science, 21(4), 043127-043127.
  
\bibitem{hemmer1959dynamic}
Hemmer, P. C. (1959). Dynamic and stochastic types of motion in the linear chain. Tapir forlag.


\bibitem{izrailev1966}
Izrailev, F. M., and Chirikov, B. V. (1966, July). Statistical properties of a nonlinear string. In Sov. Phys. Dokl (Vol. 11, No. 1, pp. 30-32).

\bibitem{ford1970stochastic}
Ford, J., and Lunsford, G. H. (1970). Stochastic behavior of resonant nearly linear oscillator systems in the limit of zero nonlinear coupling. Physical Review A, 1(1), 59.

\bibitem{livi1985equipartition}
Livi, R., Pettini, M., Ruffo, S., Sparpaglione, M., and Vulpiani, A. (1985). Equipartition threshold in nonlinear large Hamiltonian systems: The Fermi-Pasta-Ulam model. Physical Review A, 31(2), 1039.

\bibitem{cretegny1998localization}
T. Cretegny, T. Dauxois, S. Ruffo, and A. Torcini, (1998). Localization and equipartition of
energy in the $\beta$-FPU chain: Chaotic breathers. Physica D: Nonlinear Phenomena, 121,
109.

\bibitem{yoshida1990construction}
Yoshida, H. (1990). Construction of higher order symplectic integrators. Physics Letters A, 150(5), 262-268.



\end{thebibliography}







\end{article}










\end{document}